\documentclass[10pt, conference]{IEEEtran}
\IEEEoverridecommandlockouts
\usepackage{cite}
\usepackage{amsmath,amssymb,amsfonts}
\usepackage{mathfixs}
\usepackage{graphicx}
\usepackage{textcomp}
\usepackage{xcolor}
\usepackage{subfigure}
\usepackage{amsmath}
\usepackage{algorithm}
\usepackage{algorithmic}
\usepackage{caption}
\usepackage{booktabs}
\usepackage{bbm}
\usepackage{dsfont}
\usepackage{mathtools}
\usepackage[symbol]{footmisc}
\usepackage[colorlinks,linkcolor=black,anchorcolor=black,citecolor=black]{hyperref}
\newcommand{\system}{UWLoRa+~}

\def\BibTeX{{\rm B\kern-.05em{\sc i\kern-.025em b}\kern-.08em
    T\kern-.1667em\lower.7ex\hbox{E}\kern-.125emX}}

\newcommand{\ignore}[1]{}

\begin{document}

\title{Combating Multi-path Interference to Improve Chirp-based Underwater Acoustic Communication}

\author{
	\IEEEauthorblockN{Wenjun Xie$^{\dag}$, Enqi Zhang$^{\ddagger}$, Lizhao You$^{\dag}$, Deqing Wang$^{\dag}$, Zhaorui Wang$^{\star}$, and Liqun Fu$^{\dag}$}
	\IEEEauthorblockA{
		$^{\dag}$School of Informatics, Xiamen University, China\\
		$^{\ddagger}$School of Ocean, Fuzhou University, China\\
		$^{\star}$School of Science and Engineering, The Chinese University of Hong Kong, Shenzhen, China\\
		Corresponding Email: lizhaoyou@xmu.edu.cn
		}

}

\maketitle

\begin{abstract}
Linear chirp-based underwater acoustic communication has been widely used due to its reliability and long-range transmission capability. However, unlike the counterpart chirp technology in wireless -- LoRa, its throughput is severely limited by the number of modulated chirps in a symbol. The fundamental challenge lies in the underwater multi-path channel, where the delayed copied of one symbol may cause inter-symbol and intra-symbol interfere. In this paper, we present UWLoRa+, a system that realizes the same chirp modulation as LoRa with higher data rate, and enhances LoRa's design to address the multi-path challenge via the following designs: a) we replace the linear chirp used by LoRa with the non-linear chirp to reduce the signal interference range and the collision probability; b) we design an  algorithm that first demodulates each path and then combines the demodulation results of detected paths; and c) we replace the Hamming codes used by LoRa with the non-binary LDPC codes to mitigate the impact of the inevitable collision.
Experiment results show that the new designs improve the bit error rate (BER) by 3x, and the packet error rate (PER) significantly, compared with the LoRa's naive design. Compared with an state-of-the-art system for decoding underwater LoRa chirp signal, UWLoRa+ improves the throughput by up to 50 times.
\end{abstract}


\section{Introduction}  \label{sec:intro}

\begin{figure*}[t]
  \centering
  \includegraphics[width=\linewidth]{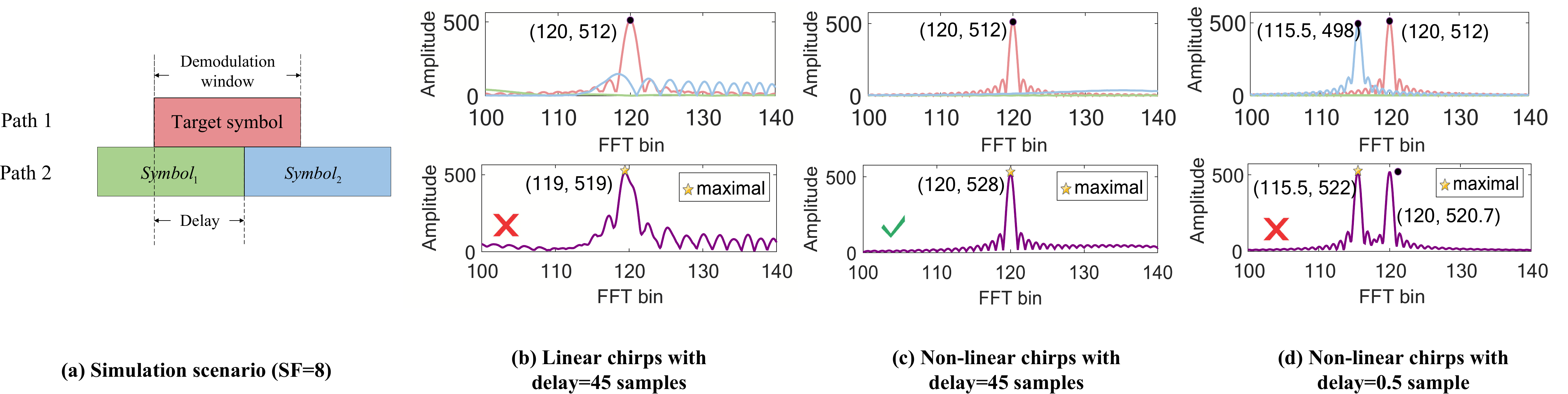}
  \caption{The peak collision examples. The above figures show their individual signals assuming the demodulation window of the target symbol. The below figures show the superimposed signals under the same window. 
  }
\label{fig:collision-example}
\end{figure*}

Underwater activities have become more and more popular in recent years, such as scuba diving or archaeological diving. One increasing demand is the reliable and high-rate communication. For example, diving people under the water are equipped with monitoring sensors, and their body status data (e.g., oxygen level) is transmitted continuously to the gateway near the sea level for safety monitoring. Video data that is captured by underwater cameras is expected to be streamed to the gateway in a real-time manner.

Chirp is a widely used technology for underwater acoustic communication (UAC). Compared with the alternative orthogonal frequency division multiplexing (OFDM) technology, the chirp technology leverages the spread spectrum technique to encode data, where the signal sweeps over a frequency range during transmission. It is more reliable and thus achieves longer transmission range, well-suited for the above applications.
There exist a few chirp-based UAC systems \cite{lei2012implementation, neasham2015development, jia2022two, steinmetz2022taking,petroni2023feasibility}. Refs. \cite{lei2012implementation, neasham2015development} adopt binary orthogonal keying (BOK), where the up-chirp signal stands for 1 and the down-chirp signal stands for 0. 
BOK is robust to the challenging acoustic channel. However, it has a low data rate, where each symbol only encodes one bit. Ref. \cite{jia2022two} uses the initial frequencies of the chirp signal to encode bits. It divide the initial frequencies into several parts, and let the gap  between any two initial frequencies be larger than the maximal path delay. In this way, they can encode more bits.

The idea of encoding more bits in a chirp symbol has been used by the Long Range (LoRa) technology \cite{liando2019known} in wireless communication. In particular, given a spreading factor (SF) number $SF$ ($SF\geq6$), each LoRa symbol encodes $SF$ bits. Although the data rate is increased, it is non-trivial to be applied to the underwater channel environment. The key challenge (more details in Section \ref{sec:challenge}) lies in that the delayed copies of the previous symbol may interfere with the current symbol (i.e., \emph{inter-symbol interference}), and the delayed copies of the current symbol may also interfere with the current symbol (i.e., \emph{intra-symbol interference}). Simply using the strongest signal for decoding as in \cite{steinmetz2022taking,petroni2023feasibility} does not work well since the introduced interference may make the strongest signal not a valid data.

In this paper, we present \system that adopts the LoRa's dense chirp modulation, and enhances LoRa's design to overcome the challenging underwater multi-path channel.
We first replace the linear chirp used by LoRa with the non-linear chirp for modulation. One advantage of the non-linear chirp is that the energy of the interference signal gets spread if the demodulation window is not aligned with the transmission window, and the inter-symbol and intra-symbol interference is greatly reduced.
Then, given the non-linear chirp design, we  treat each path independently, and align the demodulation window to each detected path for demodulation.

Although the signal collision probability is reduced, signal collision may still be inevitable. To improve the demodulation performance, we design a multi-path combination mechanism to merge demodulation results of each detected path. The intuition is that some detected paths may be less interfered and can be fully utilized by combination. 
Another design is to introduce advanced non-binary low-density parity-check (NB-LDPC) codes to combat the inevitable demodulation errors. Compared with the binary Hamming codes used by LoRa that have a short block length, NB-LDPC not only matches the M-ary chirp modulation, but also has a longer block length, more reliable to the demodulation errors.

We perform experiments to collect real chirp signals in an indoor pool, and evaluate the performance of our proposed UWLoRa+. Experiment results show that 1) the non-linear chirp reduces the symbol error rate (SER) by 2 times and the bit error rate (BER) by 3 times compared with the linear chirp for SNR$\geq$-10dB; 2) the adoption of NB-LDPC makes PER=0 for SNR$\geq$-5dB, while Hamming codes have PER=1 even for high SNRs (e.g., 10dB); and 3) the use of multi-path combination combination outperforms the straightforward algorithm that adopts the strongest path for decoding by 0.3 times to 50 times in terms of the throughput.

Overall, this paper makes the following contributions:
\begin{enumerate}
    \item We identify the multi-path interference problem, and adopt the non-linear chirp and the multi-path combination methods for improved demodulation performance.
    \item We adopt NB-LDPC to match the M-ary chirp modulation for improved decoding performance, and overcome the inevitable interference problem.
    \item We perform extensive experiments in an indoor pool to evaluate our new designs, and show the throughput improvement over the state-of-the-art system.
\end{enumerate}


\section{Background and Challenge}
This section presents the system model as background and illustrates the challenge to be tackled.
\subsection{System Model}

\textbf{Transmitted Signal.} Linear frequency modulation (LFM) is the most commonly used chirp signal. The frequency of the LFM signal linearly increase (i.e., upchirp) or decrease (i.e., downchirp). Let the system bandwidth be $BW$. The frequency of the upchirp ranges from $-\frac{BW}{2}$ to $\frac{BW}{2}$. Spreading factor (SF) defines the gradient of the frequency sweeping. 
A chirp symbol contains $N=2^{SF}$ samples with duration $T_s = \frac{2^{SF}}{BW}$.

The BOK system \cite{lei2012implementation,neasham2015development} simply encodes bit 0 with the upchirp, and bit 1 with the downchirp. Other systems \cite{jia2022two, steinmetz2022taking,petroni2023feasibility,liando2019known} only use upchirp, and use the initial frequency to encode data bits. Let the base upchirp signal with initial frequency $-\frac{BW}{2}$ be written as $C(t)=e^{j2\pi(\frac{BW}{2 T_s}t-\frac{BW}{2})t}$. 
We can cyclically shift the base upchirp to encode more bits. In particular, we can encode $n$ bits into signal $x(t)$ with an initial frequency shift $f(s)=s \cdot \frac{BW}{2^n}$, $s \in [0, 2^n)$, on $C(t)$. The nCSK system \cite{jia2022two} choose $n$ to tolerate the maximal path delay.  LoRa \cite{liando2019known} chooses $n=SF$ without considering multi-path.

Then, given a signal with frequency shift $f(s)$, the symbol frequency starts from $f(s) - \frac{BW}{2}$, increases linearly with time until it reaches $\frac{BW}{2}$ at $t=t_{fold}$, and folds down to $-\frac{BW}{2}$. That is, the transmitted signal of data symbol $s$ is given by
\begin{equation}
	\small
	\begin{aligned}
		x(t)&=
		\begin{cases}
			e^{j2\pi(\frac{BW}{2 T_s}t+f(s)-\frac{BW}{2})t},\ 0\leq t\leq t_{fold}, \\
			e^{j2\pi(\frac{BW}{2 T_s}t+f(s)-\frac{3BW}{2})t},\ t_{fold}\textless t\leq T_s. 
		\end{cases} \\
		\label{eq:x_t}
	\end{aligned}
\end{equation}

Linear chirps sweep the frequency linearly. Non-linear chirps sweep the frequency using a non-linear function, e.g., polynomial, logarithmic, or exponential. For example, for the quadratic non-linear chirp, we can write the transmitted signal of data symbol $s$ as
\begin{equation}
	\small
	\begin{aligned}
		x(t)&=
		\begin{cases}
			e^{j2\pi(\frac{BW}{ {3(T_s)}^2}t+f(s)-\frac{BW}{2})t},\ 0\leq t\leq t_{fold}, \\
			e^{j2\pi(\frac{BW}{{3(T_s)}^2}t+f(s)-\frac{3BW}{2})t},\ t_{fold}\textless t\leq T_s. 
		\end{cases} \\
		\label{eq:x_t}
	\end{aligned}
\end{equation}

\textbf{Received Signal.}
The underwater environment has severe multi-paths. For example, due to the slow sound speed under the water (i.e., around 1500m/s), a path difference of 2.5m for 6KHz signal generates a noticeable new path. 
Let $h$ be the channel, and $y(t)$ be the received signal of $x(t)$. Here $h$ is a vector, denoting a multi-path channel. 
In addition to the distortion by channel, a carrier frequency offset (CFO) $\delta$ between the transmitter and the receiver rotates the phase of the received signal. Thus, the received signal can be written as 
\begin{equation}
    y(t) = (h * x(t)) \cdot e^{j2\pi \delta t} + w(t),
\end{equation}
where $*$ means convolution, and $w$ is the additive white Gaussian noise (AWGN) noise. 
Let $N_p$ be the number of paths. The received signal can be re-written as
\begin{equation}
    y(t) = \sum_{p=1}^{N_p} h_p \cdot x(t-\tau_p) \cdot e^{j2\pi \delta t} + w(t),
\end{equation}
where $h_p$ is the amplitude of the $p$-th path, and $\tau_p$ is the delay (time offset, TO) of the $p$-th path. 
Here $\tau_p$ may not be an integer multiple of $1/BW$. Thus, oversampling is commonly used to increase the timing granularity of the receiver. 
In addition, since the carrier frequency of the acoustic signal is usually low (e.g., 20KHz-30KHz), the CFO $\delta$ is small and can be ignored in practice.

\ignore{
\begin{figure}[t]
\setlength{\belowcaptionskip}{0.5cm}
\centering
\subfigure[]{%
\includegraphics[width=0.45\columnwidth]{figures/fig_stft1.pdf}
\label{fig_stft1}}
\quad
\subfigure[]{%
\includegraphics[width=0.45\columnwidth]{figures/fig_stft2.pdf}
\label{fig_stft2}}
\caption{An example of LFM-modulated signals, and the spectrogram of received symbols using frequency-domain demodulation: (a) base up-chirp, and (b) shifted data chirp symbol.}
\label{fig_stft}
\end{figure} 
}

\textbf{Demodulation.}
For the single-path channel, there are two common non-coherent approaches to demodulate $y(t)$. 
The first approach is to perform \emph{match filtering} with respect to the transmission signals of data $\{f(0), \cdots, f(2^n-1)\}$, and decide the element (and the corresponding data) that has the maximal correlation result. 
The other approach named \emph{dechirp} is to multiple $y(t)$ with the base downchirp $C^*(t)$, and then apply the discrete Fast Fourier Transform (FFT) to find the position that has the maximal value. The found peak position corresponds to the encoded data. 
Compared with the first approach, the second approach has lower complexity, and is the standard approach in demodulating the LoRa signal. 



\subsection{Challenge} \label{sec:challenge}
Although the above dechirp approach is effective for single-path systems, the approach faces the \emph{peak collision} problem when several paths exist in the system. In particular, given a demodulation window, 
each path has a peak that is determined by the modulated data and the TO with respect to the window. 
The delayed copies of the previous symbol may collide with the current symbol if the positions of these peaks are close (i.e., inter-symbol interference). 
The delayed copies of the current symbol may collide with the current symbol if the positions of these peaks are close (i.e., intra-symbol interference). 
Both may lead to a wrong decision.


Fig. \ref{fig:collision-example} shows the peak collision examples.
We consider a typical collision scenario where the current (target) symbol is interfered by the previous symbol ($symbol_1$) and the delayed copy of the current symbol ($symbol_2$) with $SF$=8 as shown in Fig. \ref{fig:collision-example}(a). We assume no noise, and the oversampling rate is 2.
The demodulation window is aligned with the target symbol, and misaligned with the two interference symbols. 
The above part of Fig. \ref{fig:collision-example}(b) shows their individual signals with path delay 45 samples, where there are one noticeable peak (i.e., the target symbol) and some small peaks (i.e., $symbol_1$ and $symbol_2$). 
However, due to the multi-path interference, the peak position of the superimposed signal changes as shown in the below part of Fig. \ref{fig:collision-example}(b), leading to a demodulation error.



\section{System Design}

This section provides two novel design in our system. Based on these design, we present the overall algorithm.

\subsection{Adopting Non-Linear Chirps} \label{design:nl}

\begin{figure}[t!]
  \centering
  \includegraphics[width=0.65\columnwidth]{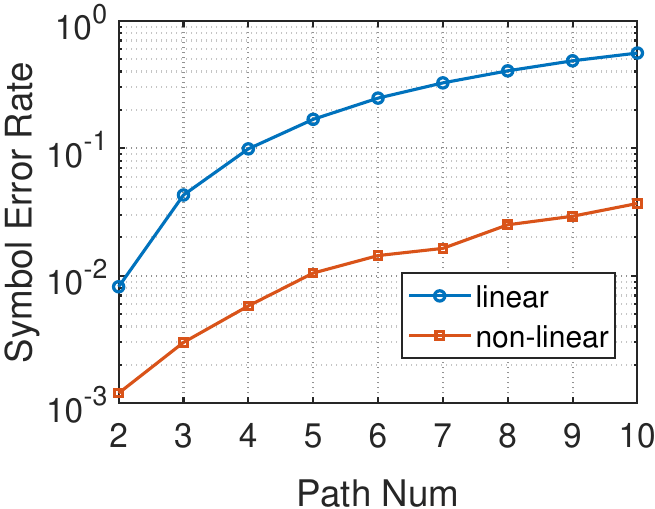}
  \caption{The symbol collision probability with the different number of paths for linear chirps and non-linear chirps.
  }
  \label{fig:collision-prob}
\end{figure}

Inspired by the algorithm \cite{li2022curvinglora} for decoding LoRa's multi-user collision signal, we first address the \emph{peak collision} problem through the non-linear chirp. 
The advantage of the non-linear chirp is the \emph{energy scattering} effect \cite{li2022curvinglora}: only when the demodulation window is almost aligned with the target signal's window, the dechirp operation shows a sharp peak; if there is a tiny misalignment, the height of peaks is greatly reduced. In contrast, for the linear chirp, a tiny misalignment only reduces the height a bit. The energy scattering effect of the non-linear chirp makes it well-suited for handling multi-path collisions, where we can shift the demodulation window to align each path without introducing much interference from other paths.
Fig. \ref{fig:collision-example}(c) shows the effect of adopting non-linear chirps under the same setup with Fig. \ref{fig:collision-example}(b). Due to the large path delay, interference symbols have negligible peaks, and do not cause a demodulation error. 

We also perform simulations to measure the peak collision probability with different number of paths. The simulation scenario is the same with Fig. \ref{fig:collision-example}(a). 
We only focus on the demodulation of the target symbol. We randomly generate the data of the target symbol and the previous symbol, and randomly generate the path delays within the range the symbol duration ($SF$=8,~$BW$=6KHz).
Fig. \ref{fig:collision-prob} shows the symbol collision probability considering the linear chirp and the non-linear chirp, where a decision error is counted as a collision. 
We can find that the symbol error rate (SER) increases dramatically as the number of paths increase, and is non-negligible (0.1) for the linear chirp with four paths. Instead, adopting non-linear chirps reduces the SER by an order of magnitude.

\subsection{Adopting Non-Binary LDPC} \label{sec:design:nb-ldpc}

Although non-linear chirps can alleviate the peak collision problem, they cannot completely remove the collision. For example, in Fig. \ref{fig:collision-example}(d), the symbol error could also happen if the path delay is small even with the non-linear chirp. We leverage channel codes to overcome the errors. One simple choice is the Hamming codes adopted by LoRa with coding rate $CR \in \{\frac{4}{5}, \frac{4}{6}, \frac{4}{7}, \frac{4}{8} \}$. Each block has at most 8 bits, and can correct at most 1 bit error. 


\begin{figure}[t!]
  \centering
  \includegraphics[width=0.65\columnwidth]{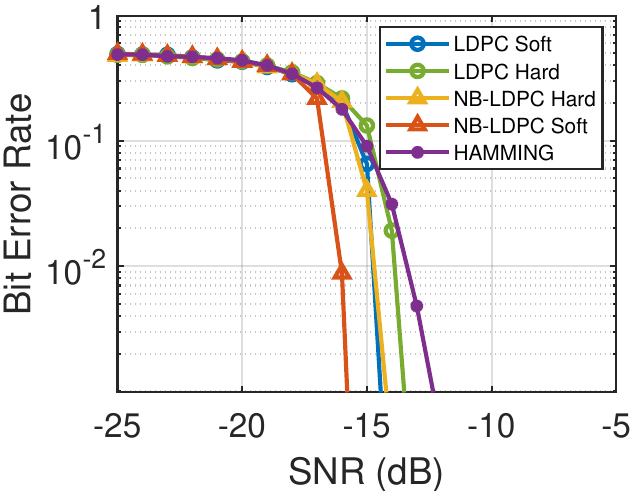}
  \caption{BER Performance of NB-LDPC, Binary-LDPC, Hamming Codes assuming the LoRa chirp modulation with $SF=8$ under the AWGN channel.}
  \label{fig:nb-ldpc-awgn}
  \vspace{-0.2cm}
\end{figure}

Although Hamming codes is very simple, its error protection capability is pretty limited. If one block of a packet cannot be corrected, the whole packet is erroneous.
When facing the error pattern in the collision symbol, the Hamming codes could not correct packet errors, since each symbol encounters the same multi-path interference and has the same error pattern. 
The experiment results in Section \ref{sec:results} show that Hamming codes have a high packet error rate under the multi-path channel. 
A new channel codes with high error protection capability and suitable implementation complexity is preferred.

We note that the advanced chirp modulation encodes several bits per symbol. It is similar to the high-order modulation such as quadrature amplitude modulation (QAM). 
\emph{Non-binary} channel codes that have a $GF(q)$ finite field can achieve better performance with
high-order modulations. Inspired by this observation, we advocate the use of non-binary channel codes in our chirp system, since a chirp symbol also encodes several bits. In particular, we adopt the non-binary LDPC codes with $GF(2^{SF})$.

One disadvantage of using the non-binary channel codes is the high decoding complexity, since there are $q$ elements whose probabilities are to be computed in the $GF(q)$ field. 
However, such problem may not be an issue for our data monitoring application, since i) there is only uplink traffic in our application, and ii) the encoding complexity (i.e., the finite field addition) can be negligible compared with the decoding complexity. Thus, the weak terminals underwater implement the encoder, and the strong gateway above the water implements the decoder.

The decoding of NB-LDPC is similar to the binary LDPC. Given the tanner graph, we use the Q-ary sum product algorithm (QSPA) \cite{barnault2003fast} for decoding. The decoding probability messages are passed from the variable nodes to the check nodes, and from the check nodes to the variable nodes, until the algorithm converges. Instead of only passing probabilities of bits $(0,1)$, QSPA passes the probabilities of bits $(0,1,\cdots,q-1)$, and the check nodes generate the update equations based on the $mod(q)$ operation. Since there are $q$ elements in the field, an update equation may involve with $q^{d-1}$ combinations for a check node with degree $d$, and thus QSPA has a high implementation complexity. We implement the Fast Fourier Transform-QSPA (FFT-QSPA) algorithm  \cite{barnault2003fast} to reduce the computation complexity.

Another issue is how to derive the Log Likelihood Ratio (LLR) probability for the NB-LDPC decoding, especially for the chirp modulation. Given the spreading factor $SF$, there are $2^{SF}$ elements in the field. We adopt the dechirp approach, and assume the amplitude of the $j$-th bin in the demodulation window is $amp_j$. The SoftMax approach is used to compute LLR. The LLR of the $j$-th bin is defined as
\begin{equation}
\begin{aligned}
    amp_{n}(j) &= \frac{amp(i)-min(amp)}{max(amp)-min(amp)},\\
    Pr(j) &= exp\left({\frac{amp_{n}(i)}{\sum amp_{n}(i)}} \right),\\
    LLR(j) &= ln\left(\frac{Pr(j)}{Pr(0)}\right), 
\end{aligned}
\end{equation}
where $max(\cdot)$ is the max function, $min(\cdot)$ is the min function, and $exp(\cdot)$ is the exponential function.

Fig. \ref{fig:nb-ldpc-awgn} shows the preliminary study of adopting NB-LDPC codes in the chirp system under the AWGN channel. We compare the performance of NB-LDPC with that of Hamming codes and binary LDPC codes, and find that NB-LDPC has the best performance under the AWGN channel, making it promising to work for the multi-path channel.


\subsection{The Overall Algorithm}

\begin{figure}[t!]
  \centering
  \includegraphics[width=\linewidth]{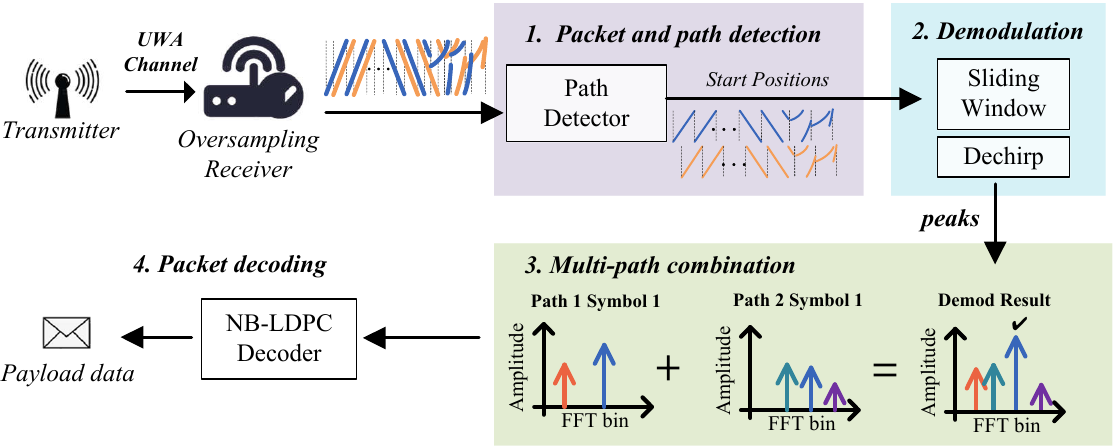}
  \caption{The overall workflow of the UWLoRa+ system.}
  \label{fig:worklow}
\end{figure}

Fig. \ref{fig:worklow} illustrates the workflow of our system which involves four key steps, i.e., packet and path detection, demodulation, multi-path combination and packet decoding. We adopt the frame format similar to LoRa with preambles, Start
Frame Delimiters (SFD), and data symbols. Preambles and SFDs are still modulated through the linear chirp, and data symbols are modulated through the non-linear chirp. Each symbol encodes $SF$ bits.

We employ an over-sampling receiver with sliding windows and cross correlation to detect the existence of a packet and all paths. In particular, we first divide the signal into chunks, and measure the energy of each chunk. If the energy surpasses the threshold, the chunk may contain preambles of the packet. Then, we take each sample within the window as a potential starting position, intercept a signal of the symbol length, and perform cross correlation with the standard preamble. Samples whose correlation results surpass the predefined threshold are considered as starting indexes of paths. Note that to alleviate the impact of sidelobes, we set a cancellation length to avoid considering samples that are within the cancellation range to be the start of the possible paths.

After detecting all possible paths, we perform demodulation for each path. In particular, we use a symbol length as the step size to sliding windows, multiply the windowed signal by the standard downchirp and then perform FFT to demodulate these time-aligned signals. Then we preserve both the frequency and amplitude of the peaks that their amplitudes exceed the threshold. The reason for not preserving the maximal peak only is that the target peak may be interfered by the multi-path signal and the sidelobes introduced by fractional TOs, and may not be the maximal one.

The preserved multiple peaks can also be used for \textbf{the multi-path combination} to leverage multi-path diversity. Note that some paths may be interfered so that the maximal peak may not be the target peak, and some paths may not be interfered so that they have a clean window with correct peaks. Therefore, we add the amplitudes of these path windows with the same data, and retain the maximal peak of the new window. In this way, interference peaks may be filtered out.

After the multi-path combination, we perform  packet decoding. To address the unavoidable demodulation errors in each symbol and match the M-ary chirp modulation, we adopt the NB-LDPC channel codes. We follow the steps in Section \ref{sec:design:nb-ldpc} to realize demodulation and decoding.


\section{Performance Evaluation} \label{sec:results}

In this section, we present the experiment setup and the experiment results.

\subsection{Experiment Setup}

\begin{figure}[t]
\setlength{\belowcaptionskip}{0.5cm}
\centering
\subfigure[]{%
\includegraphics[width=0.45\columnwidth]{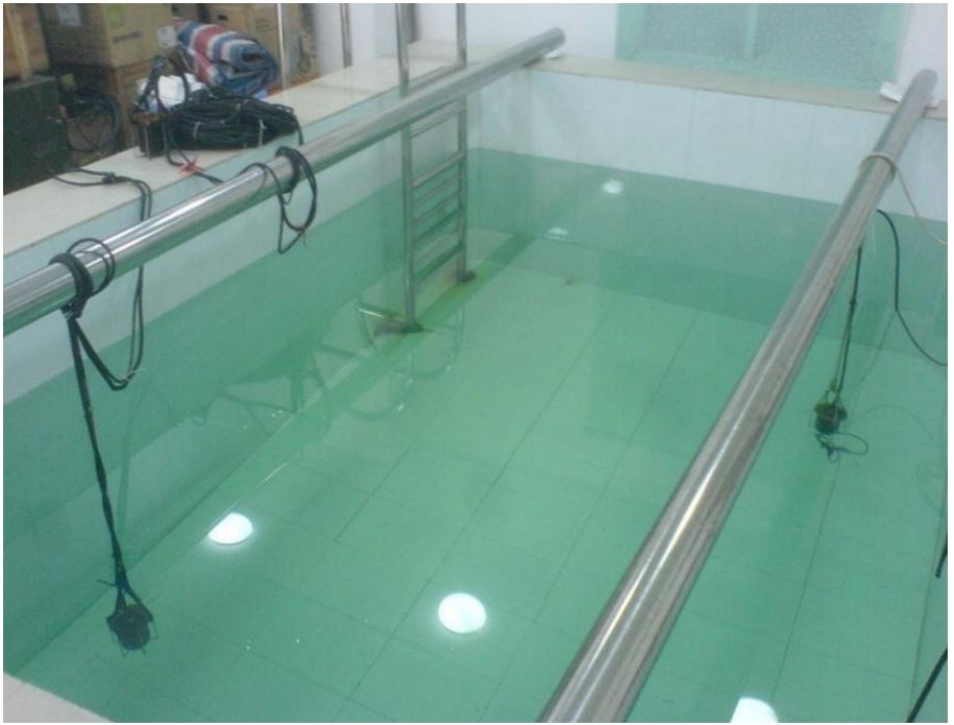}
\label{fig_env_pool}}
\quad
\subfigure[]{%
\includegraphics[width=0.45\columnwidth]{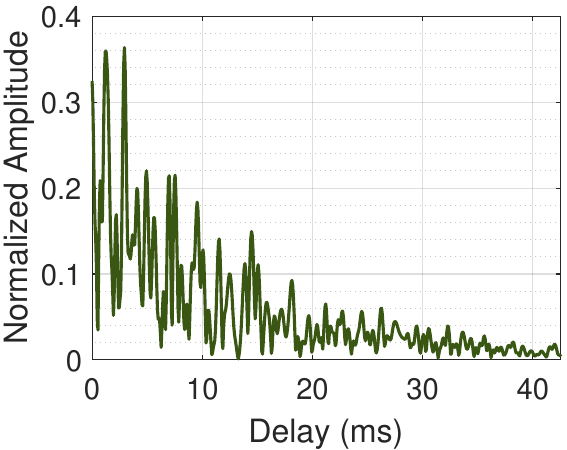}
\label{fig_env_channel}}
\caption{(a) The experiment environment: an indoor pool; (b) a typical channel measured in the pool environment.}
\label{fig_env}
\vspace{-0.6cm}
\end{figure} 

Fig. \ref{fig_env_pool} shows the experiment environment -- an indoor pool full of water with length 4.6m, width 3.0m, and depth 1.5m. We deploy two transducers, one transmitter and one receiver. Both are below the water 0.5m and have distance at least 3m. They can transmit and receive passband signals. The frequency range of the generated chirp signal is 22KHz to 28KHz with bandwidth 6KHz. The pool environment has the most challenging multi-paths due to wall reflections. Fig. \ref{fig_env_channel} shows a typical channel of the environment measured using a known signal: there are many noticeable paths, and the amplitude of other paths may be larger than the first path. UWLoRa+ adopts the quadratic non-linear chirp, $SF=8$ and NB-LDPC with the coding rate $1/2$. The source packet has 400 bits (i.e., 50 symbols). After channel coding, the packet length is 800 bits (i.e., 100 symbols).

Different from \cite{steinmetz2022taking,petroni2023feasibility} that use profiled channel for simulation, we adopt the trace-driven approach to evaluate algorithms. In particular, we generate passband signals of these packets, fix some positions, let the transmitter transmit several packets, and let the receiver log the signal. 
The processing is on the logged signal.
In each position, we randomly generate 100 packets. We measure 10 positions to sample different multi-path channel, and add the Gaussian noise to emulate different SNRs. We show the average results.



\subsection{Experiment Results}

\begin{figure*}[t]
\setlength{\belowcaptionskip}{0.5cm}
\centering
\subfigure[]{%
\includegraphics[width=0.63\columnwidth]{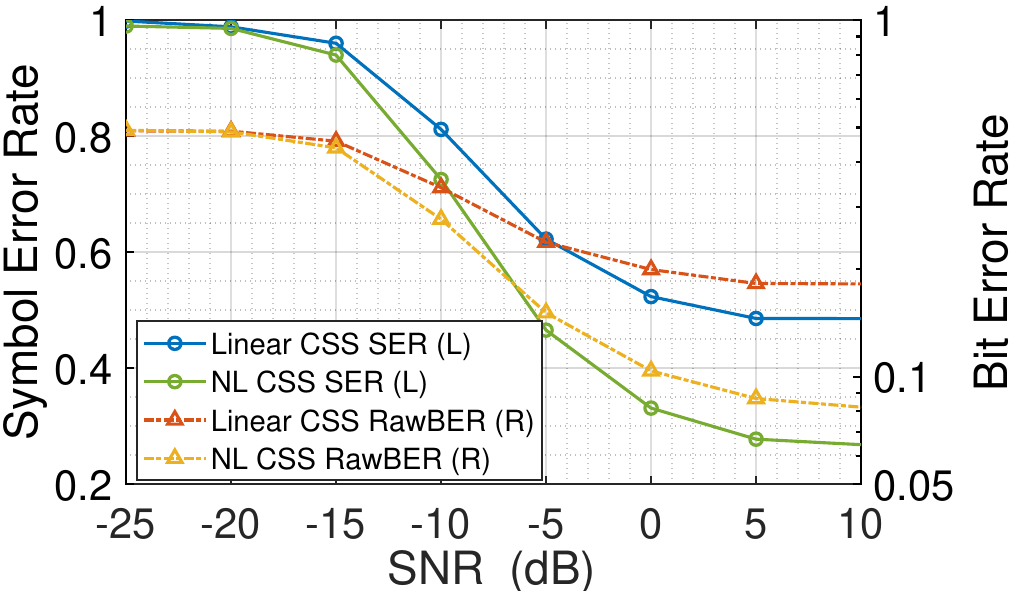}
\label{fig_ser_linear_nl}}
\quad
\subfigure[]{%
\includegraphics[width=0.61\columnwidth]{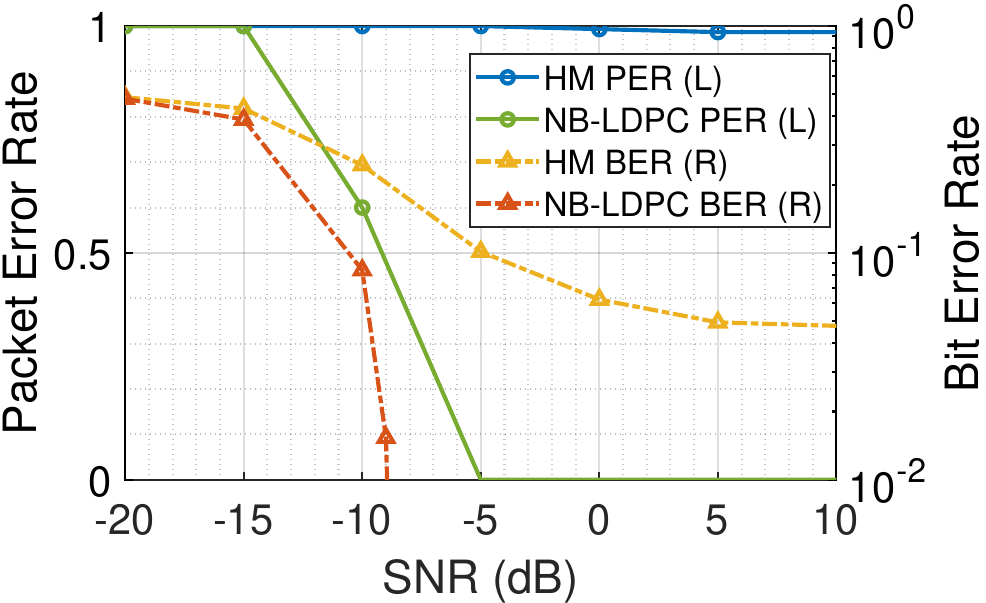}
\label{fig_per_channel_codes}}
\subfigure[]{%
\includegraphics[width=0.55\columnwidth]{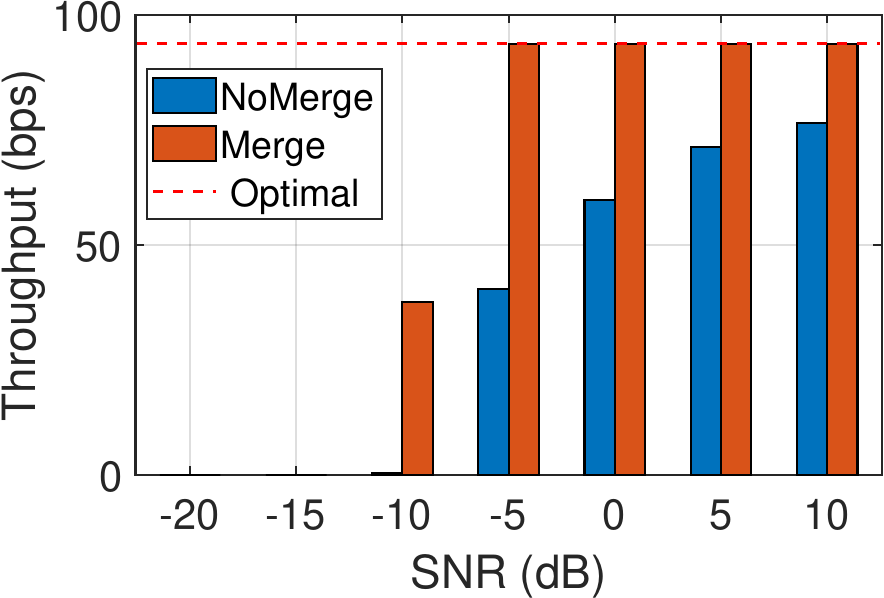}
\label{fig_throughput}}
\caption{(a) SER and RawBER results for adopting linear chirps and non-linear chirps; (b) PER and BER results for adopting Hamming codes and NB-LDPC codes; (c) throughput results of UWLoRa+ (Merge) and the SOTA system (NoMerge). }
\label{fig_env}
\vspace{-0.8cm}
\end{figure*}

\textbf{Linear vs. Non-Linear.} 
We first compare the demodulation performance using linear chirps and non-linear chirps. In the same position, we try the linear chirps and the non-linear chirps using the same data. Both have the multi-path combination.
Fig. \ref{fig_ser_linear_nl} shows the symbol error rate (SER) and bit error rate (BER) results (before channel decoding, or RawBER) for these two schemes. Non-linear chirps outperform linear chirps in both SER and BER. The improvement is 2x and 3x for SNR$\geq$-10dB, respectively. These results confirm the benefit of using non-linear chirps to reduce peak collision. Thus, we adopt the non-linear chirp in the following experiments.

\textbf{Hamming vs. NB-LDPC.} Although non-linear chirps are adopted, a BER error floor of $10^{-1}$ still exists as shown in Fig. \ref{fig_ser_linear_nl}. Channel codes are essential to remove the error floor. We compare Hamming codes and NB-LDPC. (4,8) Hamming codes are implemented. The generator matrix of NB-LDPC is generated as follows: we first generate the matrix for binary LDPC, and then replace positions that have 1 with random values within $GF(2^8)$. Since the decoding results depend on the interaction between modulated data and the channel, we apply the same demodulation error pattern to compare their performance instead of using different error patterns.

Fig. \ref{fig_per_channel_codes} shows the packet error rate (PER) after applying Hamming codes and NB-LDPC codes. The Hamming codes with block 8 cannot deal with the bit errors, and thus has PER=1 for almost all SNRs. Instead, the NB-LDPC codes have superior performance: PER becomes 0 when SNR$\geq$-5 dB. The results imply the NB-LDPC codes are suitable not only for the M-ary chirp modulation, but also for the multi-path interference scenario.

\textbf{UWLoRa+ vs. SOTA.}  Then, we compare the throughput of UWLoRa+ (i.e., \emph{Merge}) with that of the state-of-the-art (SOTA) system \cite{steinmetz2022taking} for decoding the underwater LoRa chirp signal. The SOTA system simply demodulates the strongest path ignoring all other paths (i.e., \emph{NoMerge}). For fair comparison, we also adopt NB-LDPC for the SOTA system. Fig. \ref{fig_throughput} shows their throughput. UWLoRa+ outperforms the SOTA system by 30\%-50\% ($\geq$0dB), 120\% (-5dB) to 5000\% (-10dB). The improvement is mainly due to the adoption of the multi-path combination mechanism. Meanwhile, UWLoRa+ approaches the optimal throughput 93.75bps for SNR$\geq$-5dB.


\section{Related Work}

\textbf{Chirp-based UAC:} Chirp signal has been used in UAC. Refs. \cite{steinmetz2018practical, coccolo2022underwater}  use CSS signal for preamble synchronization. Refs. \cite{lei2012implementation} and \cite{neasham2015development} use the BOK modulation. Ref. \cite{jia2022two} extends the modulation to chirp-$n$CSK, where $n$ is chosen to tolerate the maximal path delay. Ref. \cite{steinmetz2022taking} is the first paper to advocate the use of LoRa-like chirps, and further analysis is performed in \cite{petroni2023feasibility}. However, the demodulation algorithm using the strongest path limits its performance. Our system improves the decoding performance via separating/merging each path and advanced channel codes. Chirp modulation is also in combined use with FSK \cite{leblanc2000improved}, and is extended to the orthogonal chirp division multiplexing (OCDM) modulation \cite{ouyang2016orthogonal}. Our system is orthogonal to them, and can be extended to support them.

\textbf{Multi-Packet Reception for LoRa:} Decoding multi-users' LoRa signal has been well studied recently. Different from the underwater multi-path channel, they assume the single-path channel, and decode each user's data. Choir\cite{eletrebychoir17} leverages the distinct frequency offsets to identify users, which does not hold in the single-user scenario. FTrack\cite{xia2019ftrack}, OCT\cite{wang2020oct}, Pyramid\cite{xu2021pyramid}, 
AlignTrack\cite{chen2021aligntrack}, CIC\cite{shahid2021cic}, PCube\cite{xia2021pcube}, and CurvingLoRa\cite{li2022curvinglora} leverage the frequency, power and phase diversity to resolve collisions. However, they all cannot handle the peak collision problem, especially the collision problem under the multi-path channel. Our previous work LoRaPDA\cite{tang2022quick} solves the peak collision problem but assumes time-synchronized transmissions.
Our system does not have such assumption, and studies  non-linear chirps and advanced channel codes to combat the peak collision problem, and designs a multi-path combining mechanism to enhance the performance.

\section{Conclusion}
In this paper, we present a new system UWLoRa+ that enables LoRa's dense chirp modulation for UAC and tackles the severe multi-path interference problem. UWLoRa+ adopts three techniques: a) the non-linear chirp reduces the collision probability significant compared with the linear chirp; b) the multi-path combination improves the demodulation performance; and c) non-binary channel codes match the M-ary chirp modulation for improved performance, and remove the error floor due to the inevitable signal collision. Experiment results confirm the benefit of these designs, and show at most 50 times improvement over the SOTA system. In the future, we will study the mobile scenario with the Doppler effect and explore the random access scenario with multiple users.




\bibliographystyle{IEEEtran}   
\end{document}